\pdfoutput=1
\documentclass{article}%
\usepackage{amsmath}
\usepackage{cite}
\usepackage{geometry}
\usepackage{amsfonts}
\usepackage{amssymb}
\usepackage{graphicx}%
\providecommand{\U}[1]{\protect\rule{.1in}{.1in}}

\begin{document}

\title{Propagation of Vortex Electron Wave Functions in a Magnetic Field}
\author{Gregg M. Gallatin\\National Institute of Standards and Technology\\Center for Nanoscale Science and Technology\\Gaithersburg, MD 20899-6203\\gregg.gallatin@nist.gov
\and Ben McMorran\\Physics Department, University of Oregon, Eugene, OR 97403-1274}
\maketitle

\begin{abstract}
The physics of coherent beams of photons carrying axial orbital angular
momentum (OAM) is well understood and such beams, sometimes known as vortex
beams, have found applications in optics and microscopy. Recently electron
beams carrying very large values of axial OAM have been generated. In the
absence of coupling to an external electromagnetic field the propagation of
such vortex electron beams is virtually identical mathematically to that of
vortex photon beams propagating in a medium with a homogeneous index of
refraction. But when coupled to an external electromagnetic field the
propagation of vortex electron beams is distinctly different from photons.
Here we use the exact path integral solution to Schrodingers equation to
examine the time evolution of an electron wave function carrying axial OAM.
Interestingly we find that the nonzero OAM wave function can be obtained from
the zero OAM wave function, in the case considered here, simply by multipling
it by an appropriate time and position dependent prefactor. Hence adding OAM
and propagating can in this case be replaced by first propagating then adding
OAM. Also, the results shown provide an explicit illustration of the fact that
the gyromagnetic ratio for OAM is unity. We also propose a novel version of
the Bohm-Aharonov effect using vortex electron beams.

\end{abstract}

\section{Introduction}

Coherent beams of photons carrying axial orbital angular momentum (OAM),
sometimes referred to as vortex beams, are well understood.\cite{ProgOpt}%
\cite{PhysToday}\cite{PRL106} and have various uses in optics and
microscopy.\cite{OptExp}\cite{NatPhys}\cite{PNAS}\cite{NanoLett} Recently
electron beams carrying very high amounts of axial OAM have been
generated\cite{Ben} and the properties of such beams have been
studied.\cite{Verbeek}\cite{Bliokh} Mathematically the propagation of a vortex
photon beam in a medium with a homogeneous index of refraction is virtually
identical to that of a freely propagating vortex electron beam. This is
obviously not the case when the electrons are propagating in an external
electromagnetic field. Here we use the exact path integral solution to examine
how an electron wave function carrying axial OAM evolves in time. We find that
the propagation of a wave function carrying nonzero axial OAM is equivalent to
the the propagation of a zero OAM wave function multiplied by an appropriate
position and time dependent prefactor. Also, the results provide an explicit
illustration of the fact the the (non-radiatively corrected) gyromagnetic
ratio for OAM is unity as it must be.\cite{Bliokh} We will see that from a
practical point of view this means that the OAM vector rotates at half the
rate of that the electron circulates in a magnetic field, i.e., at half the
cyclotron or Landau frequency

The paper is organized as follows Section 2 briefly reviews the derivation of
the gyromagnetic ratios for orbital and spin angular momentum from the Dirac
equation Section 3 discusses the path integral solution for the
(non-relativistic) propagation of the electron wave function in a magnetic
field. Section 4 uses the path integral solution to study how a vortex
electron beam, actually a wave packet, evolves in a magnetic and shows
explicitly that the gyromagnetic ratio for OAM is unity.

\section{Dirac to Schrodinger}

For completeness we provide a brief review of the derivation of the
Schrodinger equation from the Dirac equation which shows explicitly that the
(non-radiatively corrected) gyromagnetic ratio for orbital angular momentum is
unity.\cite{Zee}

The Dirac equation in SI units is%
\begin{equation}
\left(  i\gamma^{\mu}D_{\mu}-mc\right)  \psi_{D}\left(  \vec{x},t\right)  =0
\label{1.1}%
\end{equation}
where $\psi_{D}$ is a four-component Dirac spinor and $D_{\mu}=\hbar
\partial_{\mu}-ieA_{\mu}.$Here $A_{\mu}$ is the four-vector potential and $e$
is the electron charge. The indices $\mu,\nu,\cdots$ take the values 0,1,2,3
which correspond to the $t,x,y,z$ directions, respectively \ $x_{0}%
=ct,x_{1}=x,x_{2}=y,x_{3}=z$. The Einstein summation convention wherein
repeated indices are summed over their appropriate range is used throughout,
e.g., $u_{\mu}v^{\mu}\equiv\sum_{\mu=0}^{3}u_{\mu}v^{\mu}.$

Multiplying Eq (1) by $\left(  i\gamma^{\mu}D_{\mu}+mc\right)  ,$ and using
\begin{align}
\gamma^{\mu}\gamma^{\nu}D_{\mu}D_{\nu}  &  =D^{\mu}D_{\mu}-i\sigma^{\mu\nu
}\frac{1}{2}\left[  D_{\mu},D_{\nu}\right] \nonumber\\
&  =D^{\mu}D_{\mu}-\frac{1}{2}e\hbar\sigma^{\mu\nu}F_{\mu\nu} \label{1.2}%
\end{align}
which follows from $\left\{  \gamma^{\mu},\gamma^{\nu}\right\}  =2\eta^{\mu
\nu}$ where $\gamma^{\mu}$ are the gamma matrices, $\eta^{\mu\nu} $is the
Minkowski metric, $\sigma^{\mu\nu}=\left(  i/2\right)  \left[  \gamma^{\mu
},\gamma^{\nu}\right]  $ and $F_{\mu\nu}=\partial_{\mu}A_{\nu}-\partial_{\nu
}A_{\mu}$ is the field strength tensor we get\cite{Zee}
\begin{equation}
\left(  D^{\mu}D_{\mu}-\frac{1}{2}e\hbar\sigma^{\mu\nu}F_{\mu\nu}+m^{2}%
c^{2}\right)  \psi_{D}\left(  \vec{x},t\right)  =0 \label{1.3}%
\end{equation}
Consider a\ constant magnetic field $B$ pointing the in the $z$ direction.
Using gauge invariance we can write $A_{0}=0,~A_{1}=-\frac{1}{2}Bx_{2}%
~,A_{2}=\frac{1}{2}Bx_{1},~A_{3}=0$ or equivalently $A_{i}=-\epsilon
_{ij3}\frac{B}{2}x_{j}=-\frac{B}{2}\epsilon_{ij}x_{j}$. Here $\epsilon_{ijk}$
and $\epsilon_{ij}s$are the totally antisymmetric Levi-Civita tensors.
$\epsilon_{ijk}$ is $+1\left(  -1\right)  $ when $i,j,k$ is an even(odd)
permutation of $1,2,3$ and is zero otherwise and $\epsilon_{ij}$ is $+1\left(
-1\right)  $ for $i,j=1,2\left(  2,1\right)  $ and is zero otherwise\cite{Zee}
Note that $\partial_{i}A_{i}=0$. We now have $F_{12}=-F_{21}=\partial_{1}%
A_{2}-\partial_{2}A_{1}=B.$ Working in the so called "weak field limit", i.e.
dropping the $\vec{A}^{2}$ term, gives
\begin{equation}
\left(  \hbar^{2}\left(  \frac{1}{c^{2}}\partial_{t}^{~2}-\partial_{i}%
^{~2}\right)  +ie\hbar B\left(  x_{1}\partial_{2}-x_{2}\partial_{1}\right)
-e\hbar\sigma^{12}B+m^{2}c^{2}\right)  \psi_{D}\left(  \vec{x},t\right)  =0
\label{1.4}%
\end{equation}
In the Dirac basis
\begin{equation}
\sigma^{ij}=\epsilon_{ijk}%
\begin{bmatrix}
\sigma^{k} & 0\\
0 & \sigma^{k}%
\end{bmatrix}
\label{1.5}%
\end{equation}
where the $\sigma^{k}$ are the Pauli matrices.\cite{Zee} In terms of
two-component spinors $\phi$ and $\chi,$~$\psi_{D}=%
\begin{bmatrix}
\phi\\
\chi
\end{bmatrix}
$ and for a slowly moving electron (in the Dirac basis) we can set $\chi=0$
and so finally
\begin{equation}
\left(  \hbar^{2}\left(  \frac{1}{c^{2}}\partial_{t}^{~2}-\partial_{i}%
^{~2}\right)  -eBL_{3}-e2BS_{3}+m^{2}c^{2}\right)  \phi\left(  \vec
{x},t\right)  =0 \label{1.6}%
\end{equation}
Here $L_{3}=-i\hbar\left(  x_{1}\partial_{2}-x_{2}\partial_{1}\right)  $ is
the orbital angular momentum and $S_{3}=\frac{\hbar}{2}\sigma^{3}$ is the spin
angular momentum, both in the $z$ direction. More generally\cite{Zee} we can
write%
\begin{equation}
\left(  \hbar^{2}\left(  \frac{1}{c^{2}}\partial_{t}^{~2}-\partial_{i}%
^{~2}\right)  -e\vec{B}\cdot\left(  \vec{L}+2\vec{S}\right)  +m^{2}%
c^{2}\right)  \phi\left(  \vec{x},t\right)  =0 \label{1.7}%
\end{equation}
for a constant $\vec{B}$ field. Thus we see that the OAM, $\vec{L},$ couples
to the magnetic field as $\vec{B}\cdot\vec{L}$ whereas the spin angular
momentum, $\vec{S},$ couples as $2\vec{B}\cdot\vec{S}$ and so the
(non-radiatively corrected) gyromagnetic ratio for orbital angular momentum
$g_{L}=1$ whereas for spin angular momentum $g_{S}=2.$ This difference has the
effect that electron helicity, i.e., the spin projected in the direction of
propagation, remains tangent to the trajectory, i.e, it rotates at the same
rate that the electron circulates in a magnetic field. We will see below that
because $g_{L}=1$ this is not the case for electron beams carrying axial OAM.
Note that the values of $g_{L}$ and $g_{S}$ are a property of the Hamiltonian
and not of the wave function. The vortex wave function studied below, which
carries nonzero axial OAM, still couples to the magnetic field with a $g_{L}$
value of unity$.$

\section{Path Integral Solution for Propagation in a Magnetic Field}

We are interested in OAM and not spin and so we will drop the spin term in
(\ref{1.7}) and let $\phi\left(  \vec{x},t\right)  $ be a single component
wave function. To reduce to the nonrelativistic case substitute
\begin{equation}
\phi\left(  \vec{x},t\right)  =e^{-imc^{2}t/\hbar}\psi\left(  \vec
{x},t\right)  \label{2.1}%
\end{equation}
with $\psi\left(  \vec{x},t\right)  $ slowly varying compared to $\exp\left[
-imc^{2}t/\hbar\right]  $ into (\ref{1.7}) and dropping the $\partial_{t}%
^{~2}\psi$ term we get the standard Schrodinger equation%
\begin{equation}
\left(  i\hbar\partial_{t}+\frac{\hbar^{2}}{2m}\vec{\partial}^{2}+e\vec
{B}\cdot\vec{L}\right)  \psi\left(  \vec{x},t\right)  =0 \label{2.2}%
\end{equation}
with $\vec{L}=-i\hbar\varepsilon_{ijk}\hat{x}_{i}x_{j}\partial_{k}$ where
$\hat{x}_{i}$ is the unit vector in the $i$ direction.

Because (\ref{2.2}) is linear and first order in the time derivative the
solution can be written in the form%
\begin{equation}
\psi\left(  \vec{x},t\right)  =\int d^{3}x^{\prime}K\left(  \vec{x},t,\vec
{x}^{\prime},t^{\prime}\right)  \psi\left(  \vec{x}^{\prime},t^{\prime
}\right)  \label{2.3}%
\end{equation}
where $K\left(  \vec{x},t,\vec{x}^{\prime},t^{\prime}\right)  $ is called the
"propagator" and the integral is nominally over all space. The fact that
\ (\ref{2.2}) is first order in time allows the propagator to be written as a
path integral\cite{Zee}\cite{FeynmanHibbs}\cite{Kleinert}, i.e.,
\begin{equation}
K\left(  \vec{x},t,\vec{x}^{\prime},t^{\prime}\right)  =\int\limits_{\left(
\vec{x}^{\prime},t^{\prime}\right)  }^{\left(  \vec{x},t\right)  }\delta
\vec{x}\left(  t\right)  \exp\left[  \frac{i}{\hbar}\int_{t_{a}}^{t_{b}%
}dt\mathcal{L}\left(  \vec{x}\left(  t\right)  ,\partial_{t}\vec{x}\left(
t\right)  ,t\right)  \right]  \label{2.4}%
\end{equation}
Here $\mathcal{L}\left(  \vec{x}\left(  t\right)  ,\partial_{t}\vec{x}\left(
t\right)  ,t\right)  $ is the classical Lagrangian corresponding to the
quantum Hamiltonian, and the integral is over all paths or trajectories which
go from $\vec{x}^{\prime}$ at time $t^{\prime}$ to $\vec{x}$ at time $t.$ The
Lagrangian corresponding to (\ref{2.2}) has the form%
\begin{equation}
\mathcal{L}\left(  \vec{x}\left(  t\right)  ,\partial_{t}\vec{x}\left(
t\right)  ,t\right)  =\frac{1}{2}m\left(  \partial_{t}\vec{x}\left(  t\right)
\right)  ^{2}-e\vec{A}\left(  \vec{x}\left(  t\right)  ,t\right)
\cdot\partial_{t}\vec{x}\left(  t\right)  \label{2.5}%
\end{equation}
where $\vec{A}$ is the vector potential with the magnetic field $\vec{B}%
=\vec{\partial}\times\vec{A}.$ Using the form for $\vec{A}$ given above we
get, for a constant magnetic field in the $z$ direction,%
\begin{equation}
\mathcal{L}\left(  \vec{x}\left(  t\right)  ,\partial_{t}\vec{x}\left(
t\right)  \right)  =\frac{m}{2}\left(  \partial_{t}\vec{x}\left(  t\right)
\right)  ^{2}+\frac{eB}{2}\epsilon_{ij}x_{i}\partial_{t}x_{j}\left(  t\right)
\label{2.6}%
\end{equation}
It should be noted that the Lagrangian in (\ref{2.5}) and (\ref{2.6}) is the
full Lagrangian, not the weak field approximation . This can be seen simply by
calculating the corresponding classical Hamiltonian which yields $H=\left(
\vec{p}-e\vec{A}\right)  ^{2}/2m$.with $\vec{p}=m\partial_{t}x\left(
t\right)  .$

The solution for the propagator with this Lagrangian is
straightforward\cite{FeynmanHibbs}\cite{Kleinert}, indeed it's given as a
problem in Feynman and Hibbs book.\cite{FeynmanHibbs2} Transform to a rotating
frame\ in the $xy$ or $1,2$ plane by writing
\begin{equation}
x_{i}=\exp\left[  \frac{eBt}{2m}\epsilon\right]  _{ij}X_{j}\ \ \ \Rightarrow
\ \ \ \ \binom{x_{1}}{x_{2}}=%
\begin{pmatrix}
\cos\left[  \frac{eBt}{2m}\right]  & \sin\left[  \frac{eBt}{2m}\right] \\
-\sin\left[  \frac{eBt}{2m}\right]  & \cos\left[  \frac{eBt}{2m}\right]
\end{pmatrix}
\binom{X_{1}}{X_{2}} \label{2.7}%
\end{equation}
In terms of the new variables the Lagrangian corresponds to free propagation
in the $z$ direction and a harmonic oscillator in the $X_{i},$ $i=1,2$
directions with radian frequency $eB/2m.$ The path integral solutions for free
propagation and for a harmonic oscillator are well known\cite{FeynmanHibbs}%
\cite{Kleinert}. Using these results and transforming back to the non-rotating
coordinates we get%
\begin{equation}
K\left(  \vec{x},t,\vec{x}^{\prime},t^{\prime}\right)  =\left(  \frac{m}{2\pi
i\hbar T}\right)  ^{3/2}\frac{\frac{\omega}{2}T}{\sin\left[  \frac{\omega}%
{2}T\right]  }\exp\left[  \frac{i}{2\hbar}\left(
\begin{array}
[c]{c}%
\frac{m\left(  z-z^{\prime}\right)  ^{2}}{T}+\frac{m\omega}{2}\cot\left[
\frac{\omega}{2}T\right]  \left(  x_{i}-x_{i}^{\prime}\right)  ^{2}\\
+m\omega\epsilon_{ij}x_{i}x_{j}^{\prime}%
\end{array}
\right)  \right]  \label{2.8}%
\end{equation}
with
\begin{equation}
\omega=\frac{eB}{m} \label{2.9}%
\end{equation}
which is the standard cyclotron frequency\cite{Kleinert} and $T\equiv
t-t^{\prime}.$ In (\ref{2.8}) the combination $\omega T$ always occurs divided
by 2 and so we should expect various aspects of the wave function to evolve at
half the rate at which the electron circulates in the magnetic field.

Note that in the limit as $\omega\rightarrow0$ the propagator in (\ref{2.8})
reduces to the free propagator%
\begin{equation}
K_{free}\left(  \vec{r}-\vec{r}^{\prime},t-t^{\prime}\right)  =\left(
\frac{m}{2\pi i\hbar\left(  t-t^{\prime}\right)  }\right)  ^{3/2}\exp\left[
\frac{im}{2\hbar}\frac{\left(  x_{i}-x_{i}^{\prime}\right)  ^{2}}{t-t^{\prime
}}\right]  \label{2.10}%
\end{equation}
which is explicitly space and time translation invariant as it should be.

\section{Evolution of a Gaussian wave function with and without OAM}

The propagator given in (\ref{2.8}) is Gaussian in form and so if we choose a
Gaussian for the wave function at $t^{\prime}=0$ it will remain Gaussian.
Also, in this case the integral in (\ref{2.3}) can be evaluated analytically.

First consider propagation perpendicular to the magnetic field. In this case
let the initial normalized wave function be a Gaussian centered at the origin
and propagating in the $x_{2}=y$ direction%
\begin{equation}
\psi_{0}\left(  \vec{r},0\right)  =\frac{1}{\sqrt{\pi\sigma^{2}\sqrt{\pi
L^{2}}}}\exp\left[  -\frac{x^{2}+z^{2}}{2\sigma^{2}}-\frac{y^{2}}{2L^{2}%
}+\frac{i}{\hbar}py\right]  \label{3.1}%
\end{equation}
where we have switched from the $x_{i}$ notation to the more convenient at
this stage $x,y,z$ notation with $\vec{r}=x\hat{x}+y\hat{y}+z\hat{z}$. This
wave function is roughly $\sigma$ in width in the $x$ and $z$ directions and
has length $L$ in the $y$ direction. If we specify the values of $\omega$ and
the radius $R$ of the classical orbit of the electron then $p=m\omega R.$ If
we take $\sigma$ and $L$ to be much larger than the nominal de Broglie
wavelength of $2\pi\hbar/p$ then we expect mininal "diffraction" effects to
occur during propagation and as shown explicitly below this is exactly the
case. This initial wave function has zero OAM about it's direction of
propagation, the $y$ direction, since%
\begin{equation}
L_{y}\psi_{0}\left(  \vec{r},0\right)  =i\hbar\left(  x\partial_{z}%
-z\partial_{x}\right)  \psi_{0}\left(  \vec{r},0\right)  =0 \label{3.2}%
\end{equation}
To generate axial OAM the so called ladder operator approach\cite{Sakurai} is
used. Consider an operator $\mathbf{A}$ with eigenstate $\left\vert
a\right\rangle $ such that $\mathbf{A}\left\vert a\right\rangle =a\left\vert
a\right\rangle .$ We now want to generate a state $\left\vert a+1\right\rangle
$ such that $\mathbf{A}\left\vert a+1\right\rangle =\left(  a+1\right)
\left\vert a+1\right\rangle .$ To do this we only need to find an operator
$\mathbf{B}$ such that $\left[  \mathbf{A},\mathbf{B}\right]  =\mathbf{B}$
since then $\mathbf{AB}\left\vert a\right\rangle =\mathbf{B}\left\vert
a\right\rangle +\mathbf{BA}\left\vert a\right\rangle =\left(  a+1\right)
\mathbf{B}\left\vert a\right\rangle $ and so the state $\mathbf{B}\left\vert
a\right\rangle =\left\vert a+1\right\rangle ,$ up to normalization and phase
factors. Noting that
\begin{equation}
\left[  L_{y}/\hbar,\left(  \partial_{x}-i\partial_{z}\right)  \right]
=\left[  i\left(  x\partial_{z}-z\partial_{x}\right)  ,\left(  \partial
_{x}-i\partial_{z}\right)  \right]  =\left(  \partial_{x}-i\partial
_{z}\right)  \label{3.3}%
\end{equation}
it follows that a state with 1 unit of axial OAM, $\psi_{1}\left(  \vec
{r},0\right)  ,$ is given (up to normalization and phase factors) by%
\begin{equation}
\psi_{1}\left(  \vec{r},0\right)  =\left(  \partial_{x}-i\partial_{z}\right)
\psi_{0}\left(  \vec{r},0\right)  =\frac{1}{\sigma^{2}}\left(  -x+iz\right)
\psi_{0}\left(  \vec{r},0\right)  =\frac{1}{\sigma^{2}}\rho e^{i\theta}%
\psi_{0}\left(  \vec{r},0\right)  \label{3.4}%
\end{equation}
Here $\rho=\sqrt{x^{2}+z^{2}}$ and $\theta$ increases in the counterclockwise
direction when looking in the $-y$ direction and is measured from the $-x$
axis. Using the fact that $i\left(  x\partial_{z}-z\partial_{x}\right)
=-i\partial_{\theta}$ we immediately see that $L_{y}\psi_{1}=\hbar\psi_{1}%
.$and so $\psi_{1}$ carries one unit of axial OAM. The factor of $\rho,$ which
appears automatically, is necessary since at $\rho=0$ (= the $y$ axis in this
case) the phase $\exp\left[  i\theta\right]  $ is not defined and the wave
function must vanish there.

Substituting $\psi_{0}\left(  \vec{r},0\right)  $ into (\ref{2.3}) and using
(\ref{2.8}) gives%
\begin{align}
\psi_{0}\left(  \vec{r},t\right)   &  =N\int d^{3}r^{\prime}\exp\left[
\begin{array}
[c]{c}%
\begin{array}
[c]{c}%
\frac{im}{2\hbar t}\left(  z-z^{\prime}\right)  ^{2}+\frac{im\omega}{4\hbar
}\cot\left[  \frac{\omega t}{2}\right]  \left(  \left(  x-x^{\prime}\right)
^{2}+\left(  y-y^{\prime}\right)  ^{2}\right) \\
+\frac{im\omega}{2\hbar}\left(  xy^{\prime}-yx^{\prime}\right)
\end{array}
\\
-\frac{1}{2\sigma^{2}}\left(  x^{\prime2}+z^{\prime2}\right)  -\frac{1}%
{2L^{2}}y^{\prime2}+\frac{im\omega R}{\hbar}y^{\prime}%
\end{array}
\right] \nonumber\\
&  =N\exp\left[  \frac{im}{2\hbar t}z^{2}+\frac{im\omega}{4\hbar}\cot\left[
\frac{\omega t}{2}\right]  \left(  x^{2}+y^{2}\right)  \right] \nonumber\\
&  \times\int d^{3}r^{\prime}\exp\left[  \alpha_{x}x^{\prime}+\alpha
_{y}y^{\prime}+\alpha_{z}z^{\prime}-\frac{1}{2\beta_{x}}x^{\prime2}-\frac
{1}{2\beta_{y}}y^{\prime2}-\frac{1}{2\beta_{z}}z^{\prime2}\right] \nonumber\\
&  =N\exp\left[  \frac{im}{2\hbar t}z^{2}+\frac{im\omega}{4\hbar}\cot\left[
\frac{\omega t}{2}\right]  \left(  x^{2}+y^{2}\right)  \right] \nonumber\\
&  \times\sqrt{\left(  2\pi\right)  ^{3}\beta_{x}\beta_{y}\beta_{z}}%
\exp\left[  \frac{1}{2}\beta_{x}\alpha_{x}^{~2}+\frac{1}{2}\beta_{y}\alpha
_{y}^{~2}+\frac{1}{2}\beta_{z}\alpha_{z}^{~2}\right]  \label{3.5}%
\end{align}
where
\begin{align}
N  &  =\left(  \frac{m}{2\pi i\hbar t}\right)  ^{3/2}\frac{\frac{\omega t}{2}%
}{\sin\left[  \frac{\omega t}{2}\right]  }\frac{1}{\sqrt{\pi\sigma^{2}%
\sqrt{\pi L^{2}}}}\nonumber\\
\alpha_{x}  &  =-\frac{im\omega}{2\hbar}\cot\left[  \frac{\omega t}{2}\right]
x-\frac{im\omega}{2\hbar}y\nonumber\\
\alpha_{y}  &  =-\frac{im\omega}{2\hbar}\cot\left[  \frac{\omega t}{2}\right]
y+\frac{im\omega}{2\hbar}x+\frac{im\omega R}{\hbar}\nonumber\\
\alpha_{z}  &  =-\frac{im}{\hbar t}z\label{3.6}\\
\beta_{x}  &  =\left(  \frac{1}{\sigma^{2}}-\frac{im\omega}{2\hbar}\cot\left[
\frac{\omega t}{2}\right]  \right)  ^{-1}\nonumber\\
\beta_{y}  &  =\left(  \frac{1}{L^{2}}-\frac{im\omega}{2\hbar}\cot\left[
\frac{\omega t}{2}\right]  \right)  ^{-1}\nonumber\\
\beta_{z}  &  =\left(  \frac{1}{\sigma^{2}}-\frac{im}{\hbar t}\right)
\nonumber
\end{align}

To propagate $\psi_{1}$ we can write%
\begin{align}
\psi_{1}\left(  \vec{r},t\right)   &  =N\int d^{3}r^{\prime}K\left(  \vec
{r},t,\vec{r}^{\prime},0\right)  \left(  \partial_{x^{\prime}}-i\partial
_{z/}\right)  \psi_{0}\left(  \vec{r}^{\prime},0\right) \nonumber\\
&  =\frac{N}{\sigma^{2}}\int d^{3}r^{\prime}K\left(  \vec{r},t,\vec{r}%
^{\prime},0\right)  \left(  -x^{\prime}+iz^{\prime}\right)  \psi_{0}\left(
\vec{r}^{\prime},0\right) \nonumber\\
&  =\frac{N}{\sigma^{2}}\left.  \partial_{\lambda}\int d^{3}r^{\prime}K\left(
\vec{r},t,\vec{r}^{\prime},0\right)  \exp\left[  \lambda\left(  -x^{\prime
}+iz^{\prime}\right)  \right]  \psi_{0}\left(  \vec{r}^{\prime},0\right)
\right\vert _{\lambda=0} \label{3.7}%
\end{align}
The integral is still Gaussian and can be evaluated as above by letting
$\alpha_{x}\rightarrow\alpha_{x}-\lambda$ and $\alpha_{z}\rightarrow\alpha
_{z}+i\lambda$ in (\ref{3.5}). Taking the derivative with respect to $\lambda$
and setting $\lambda=0$ then yields%
\begin{align}
\psi_{1}\left(  \vec{r},t\right)   &  =\frac{N}{\sigma^{2}}\exp\left[
\frac{im}{2\hbar t}z^{2}+\frac{im\omega}{4\hbar}\cot\left[  \frac{\omega t}%
{2}\right]  \left(  x^{2}+y^{2}\right)  \right] \nonumber\\
&  \times\sqrt{\left(  2\pi\right)  ^{3}\beta_{x}\beta_{y}\beta_{z}}\left(
-\beta_{x}\alpha_{x}+i\beta_{z}\alpha_{z}\right)  \exp\left[  \frac{1}{2}%
\beta_{x}\alpha_{x}^{~2}+\frac{1}{2}\beta_{y}\alpha_{y}^{~2}+\frac{1}{2}%
\beta_{z}\alpha_{z}^{~2}\right] \nonumber\\
&  =\left(  -\beta_{x}\alpha_{x}+i\beta_{z}\alpha_{z}\right)  \frac{1}%
{\sigma^{2}}\psi_{0}\left(  \vec{r},t\right)  \label{3.8}%
\end{align}
with $\alpha_{x},\beta_{x},\ldots$the same as in (\ref{3.6}).

Even though both these analytic solutions can be manipulated into somewhat
more convenient forms, this is not very illuminating and so we will simply
plot these solutions for a set of conditions which nicely illlustrate the
relevant aspects of their time evolution. On the other hand it is worthwhile
to examine the factor $\left(  -\beta_{x}\alpha_{x}+i\beta_{z}\alpha
_{z}\right)  $ to get a better understanding of how it evolves and controls
the orientation of the OAM. Substituting from above we find, after some
algebra,%
\begin{equation}
f\left(  \vec{r},t\right)  \equiv-\beta_{x}\alpha_{x}+i\beta_{z}\alpha
_{z}=\frac{\cos\left[  \frac{\omega t}{2}\right]  x+\sin\left[  \frac{\omega
t}{2}\right]  y}{\left(  \sin\left[  \frac{\omega t}{2}\right]  \frac{2\hbar
}{im\omega\sigma^{2}}-\cos\left[  \frac{\omega t}{2}\right]  \right)  }%
+i\frac{z}{\left(  1-\frac{\hbar t}{im\sigma^{2}}\right)  } \label{3.9}%
\end{equation}
We see that $f\left(  \vec{r},0\right)  =-x+iz$ at $t=0,$ as it should, and
that it rotates in time in the $xy$ plane at a radian frequency of $\omega/2,$
The origin of this factor obvious. In operator notation, ignoring the
$1/\sigma^{2}$, (\ref{3.4}) becomes
\begin{equation}
\left\vert \psi_{1}\right\rangle =\left(  -\mathbf{X}+i\mathbf{Z}\right)
\left\vert \psi_{0}\right\rangle \label{3.10}%
\end{equation}
The time evolution is given by
\begin{align}
e^{-i\mathbf{H}t/\hbar}\left\vert \psi_{1}\right\rangle  &  =e^{-i\mathbf{H}%
t/\hbar}\left(  -\mathbf{X}+i\mathbf{Z}\right)  \left\vert \psi_{0}%
\right\rangle \nonumber\\
&  =\left(  e^{-i\mathbf{H}t/\hbar}\left(  -\mathbf{X}+i\mathbf{Z}\right)
e^{+i\mathbf{H}t/\hbar}\right)  e^{-i\mathbf{H}t/\hbar}\left\vert \psi
_{0}\right\rangle \nonumber\\
&  =f\left(  \overset{\rightarrow}{\mathbf{R}},t\right)  e^{-i\mathbf{H}%
t/\hbar}\left\vert \psi_{0}\right\rangle \label{3.11}%
\end{align}
where $\mathbf{H=}\left(  \overset{\rightarrow}{\mathbf{P}}-e\vec{A}\left(
\overset{\rightarrow}{\mathbf{R}}\right)  \right)  ^{2}/2m$ is the quantum
Hamiltonian corresponding to the Lagrangian (\ref{2.6}). Note this is the full
Hamiltonian, not the weak field approximation.

The position of the node of $\psi_{1}\left(  \vec{r},t\right)  $ follows from
the solution to $f\left(  \vec{r},t\right)  =0.$ At $t=0$ this is the $y$ axis
as shown above. For arbitrary $t$ we have the solution%
\begin{align}
y &  =-\cot\left[  \frac{\omega t}{2}\right]  x\nonumber\\
z &  =0\label{3.12}%
\end{align}
This solution is illustrated in Figure 1 for several values of $t$. This
"nodal line" rotates only by $\pi$ during one full period, $\tau=2\pi/\omega,$
of the electron cyclotron orbit and since this factor is the origin of the OAM
carried by $\psi_{1}$ this shows explicity that the OAM rotates at half the
cyclotron frequency, i.e., $g_{L}=1.$ This also shows that the OAM is axially
oriented only at times $t=n\tau,$ with $n=0,1,2,\cdots$, and its direction
switches between being parallel and antiparallel to the direction of
propagation at each of these times.%
\begin{figure}[ptb]%
\centering
\includegraphics[
natheight=3.900300in,
natwidth=4.166700in,
height=3.5371in,
width=3.7775in
]%
{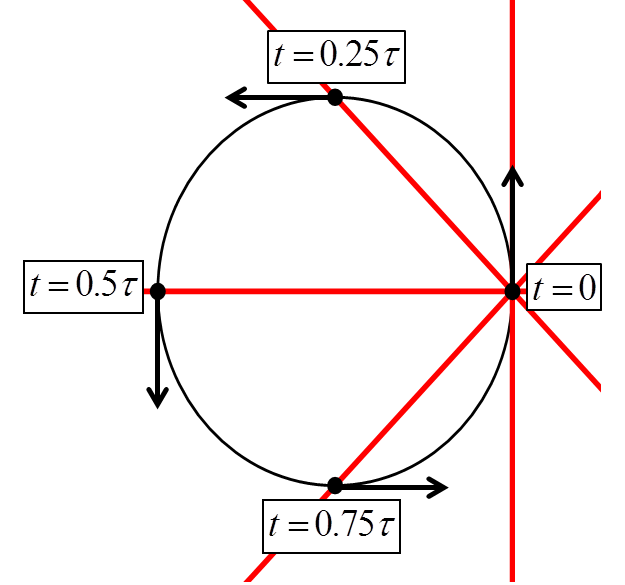}%
\caption{The graph shows the nodal lines (red) at different positions in the
electron orbit. The OAM lies along the nodal lines and thus rotates at half
the cyclotron frequency $\omega=eB/m.$}%
\end{figure}

Note that $\psi_{0}\left(  \vec{r},t\right)  $ and $\psi_{1}\left(  \vec
{r},t\right)  $ are not simply propagating Gaussian envelope functions
multiplied by a propagating plane wave factor of the form $\exp\left[
i\vec{p}\cdot\vec{r}/\hbar-iEt/\hbar\right]  $ with $\left\vert \vec
{p}\right\vert $ constant (but rotating at radian frequency $\omega)$ and
$E=\left\vert \vec{p}\right\vert ^{2}/2m$. For both wave functions the de
Broglie wavelength varies in time. This is to be expected since the coupling
to the vector potential contributes an extra phase to the wave function of the
form $-i/\hbar\int_{0}^{t}dt\vec{A}\left(  \vec{r}\right)  \cdot\partial
_{t}\vec{r}\left(  t\right)  $ which varies with position in generally an
nonlinear fashion . Figures 2 and 3 show slices of the modulus squared and the
real parts of $\psi_{0}$ and $\psi_{1}$ in the $xy$ plane at different
positions in the electron orbit. The values chosen for $\sigma,L,\omega$ and
$R$ are such that the size of the wave packet at $t=0$, $L$ in the $y$
direction and $\sigma$ in the $x$ direction are both much larger than the
wavelength (so that diffraction effects are minimal) and $R$ is much larger
than $L$. The actual ratios used for the plots are $R=10^{3}L,~L=10\sigma$ and
$\sigma\simeq10^{5}2\pi\hbar/m\omega$ hence the spatial range of the
$\operatorname{Re}\left[  \psi_{0}\right]  $ and $\operatorname{Re}\left[
\psi_{1}\right]  $ plots is about 5 orders of magnitude smaller than for the
$\left\vert \psi_{0}\right\vert ^{2}$ and $\left\vert \psi_{1}^{2}\right\vert
$ plots so that the phase variation is visible. In Figure 2 we see that the
long axis of the wave function tracks the nodal line and the spatial extent of
the wave function varies with period $\tau$ and thus the length and width
return, up to diffraction effects to their initial values at every
$t=\tau,~2\tau,~3\tau,\cdots.$ This periodic variation in the spatial extent
of the wave function can be traced back to the fact that in the rotating frame
the Lagrangian is that of a harmonic oscillator.The free propagation part of
the Langrangian, $m\left(  \partial_{t}x\right)  ^{2}/2$ cause the wave
function to expand or diffract as it propagates. The harmonic oscillator part,
$m\omega^{2}\vec{x}^{2}/2$ causes the wave function to contract and unless
these two effects are precisely balanced the wave function will oscillate in
size This is exactly analogous to the propagation of a paraxial Gaussian
optical beam.centered on the $z$ axis and propagating in the $z$ direction in
a medium with an index of refraction of the form $n\left(  x,y\right)
=n_{0}-c\left(  x^{2}+y^{2}\right)  $, i.e, a harmonic osciallator potential.
In the paraxial approximation the propagator for the photon beam has the same
Gaussian form as the propagator for the harmonic oscillator. The quadratic
variation of the index of refraction will case the beam to focus or shrink in
size as it propagates whereas diffraction effects cause the beam to expand as
it propagates. If the beam is large, so that the focusing effect dominates,
then the beam will shrink in size as it propagates. Eventually it reaches a
size where the diffraction effect dominates and it begins to expand. This
process repeats itself causing the beam to oscillate in size with a fixed
period along its length.\cite{YarivYeh} These oscillations can be prevented if
the size of the beam is fine tuned so that the diffraction and focusing
effects exactly cancel out.\cite{YarivYeh} Figure 3 shows the propagation of
the wave function $\psi_{1}$ carrying a single unit of OAM. The node in the
center of the wave function maintains its alignment on the nodal line during
each cycle. The spiral form the phase of $\psi_{1}$ is apparent in the
$\operatorname{Re}\left[  \psi_{1}\right]  $ plots. Clearly the OAM is
rotating at half the cyclotron frequency $\omega$.%
\begin{figure}[ptb]%
\centering
\includegraphics[
natheight=6.000100in,
natwidth=7.253200in,
height=4.2281in,
width=5.1059in
]%
{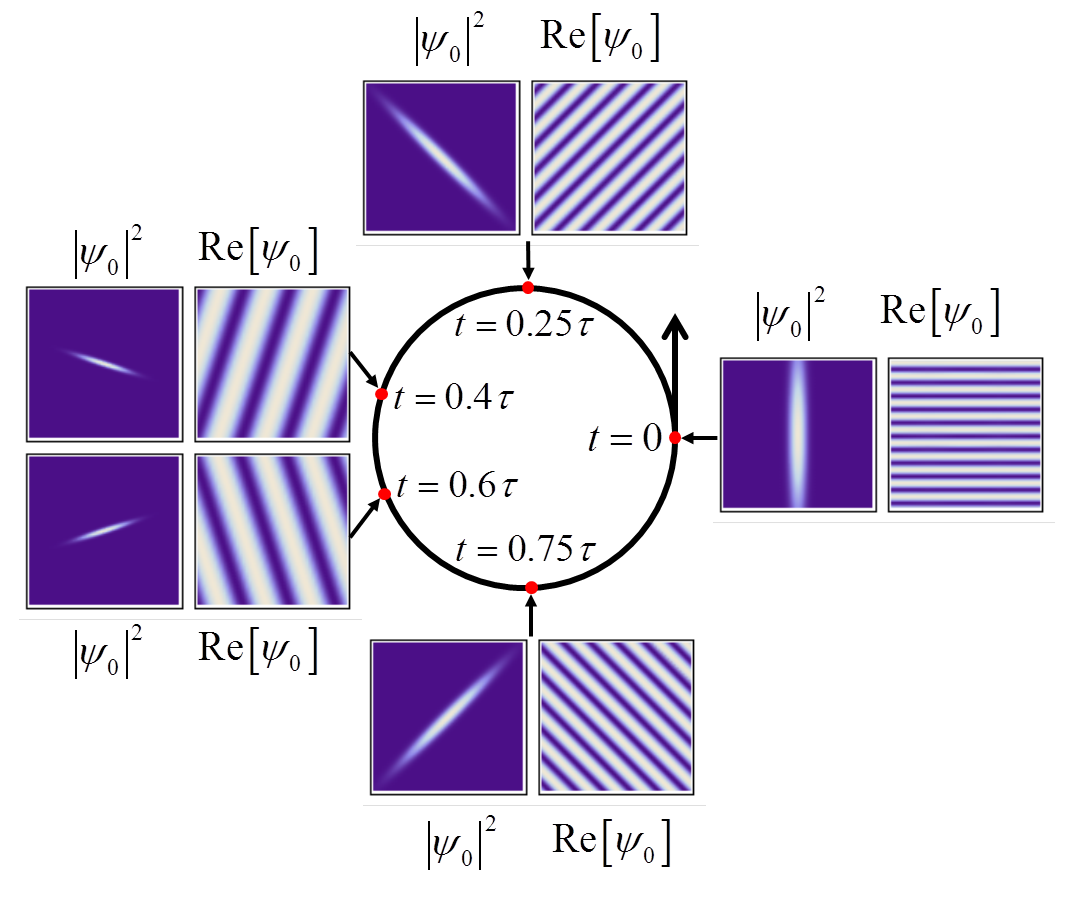}%
\caption{Slices in the $xy$ plane of $\left\vert \psi_{0}\right\vert ^{2}$ and
$\operatorname{Re}\left[  \psi_{0}\right]  $ at different positions around the
cyclotron orbit where $\psi_{0}$ is a Gaussian wavepacket carrying 0 axial
orbital angular momentum(OAM). The values chosen for the width $\sigma$ and
length $L$ of the wavepacket$,$ the cyclotron frequency $\omega=eB/m,$ and the
radius of the cycloctron orbit $R$ are such that the size of the wave packet
at $t=0$ ($L$ in the $y$ direction and $\sigma$ in the $x$ direction) are much
larger than the wavelength so that diffraction effects are minimal. All the
plots are the same fixed spatial scale with that of the $\operatorname{Re}%
\left[  \psi_{0}\right]  $ plots being about 5 orders of magnitude smaller
than the $\left\vert \psi_{0}\right\vert ^{2}$ plots so that the phase of the
wavepacket is visible. At $t=0.5\tau$ the wavepacket would be too small to be
seen at this fixed spatial scale and so it is shown at times $t=0.4\tau$ and
$t=0.6\tau$ instead. }%
\end{figure}

\begin{center}%
\begin{figure}[ptb]%
\centering
\includegraphics[
natheight=6.000100in,
natwidth=7.246300in,
height=4.2272in,
width=5.1007in
]%
{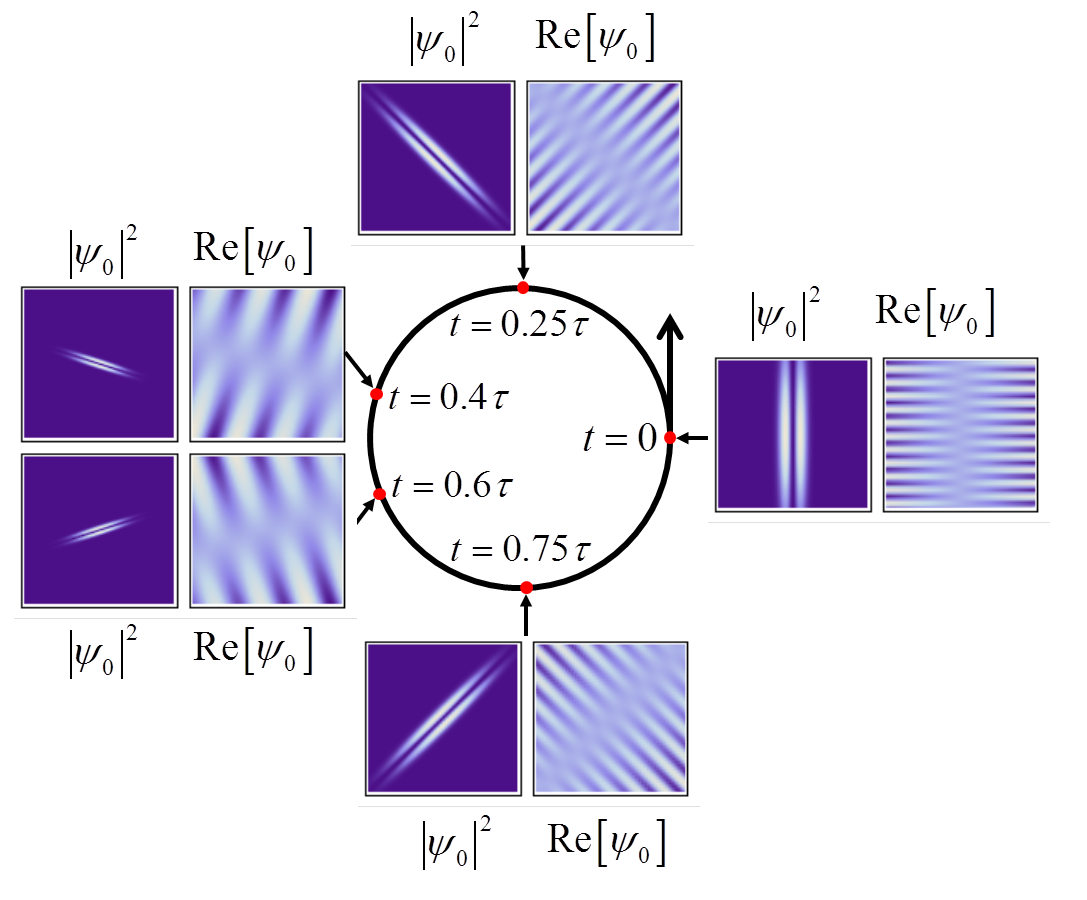}%
\caption{Slices in the $xy$ plane of $\left\vert \psi_{1}\right\vert ^{2}$ and
$\operatorname{Re}\left[  \psi_{1}\right]  $ at different positions around the
cyclotron orbit where $\psi_{1}$ is a Gaussian wavepacket carrying 1 unit
axial orbital angular momentum(OAM) oriented in the $y$ direction at $t=0$.
The values chosen for the width $\sigma$ and length $L$ of the wavepacket$,$
the cyclotron frequency $\omega=eB/m,$ and the radius of the cycloctron orbit
$R$ are the same as in Figure 2, i.e., they are such that the size of the wave
packet at $t=0$ ($L$ in the $y$ direction and $\sigma$ in the $x$ direction)
are much larger than the wavelength so that diffraction effects are minimal.
All the plots are the same fixed spatial scale with that of the
$\operatorname{Re}\left[  \psi_{1}\right]  $ plots being about 5 orders of
magnitude smaller than the $\left\vert \psi_{1}\right\vert ^{2}$ plots so that
the phase of the wavepacket is visible. At $t=0.5\tau$ the wavepacket would be
too small to be seen at this fixed spatial scale and so it is shown at times
$t=0.4\tau$ and $t=0.6\tau$ instead. }%
\end{figure}

\end{center}

Now consider propagation parallel to the magetic field. In this case we let
\begin{equation}
\psi_{0}\left(  \vec{r},0\right)  =\frac{1}{\sqrt{\pi\sigma^{2}\sqrt{\pi
L^{2}}}}\exp\left[  -\frac{x^{2}+y^{2}}{2\sigma^{2}}-\frac{z^{2}}{2L^{2}%
}+\frac{i}{\hbar}pz\right]  \label{3.13}%
\end{equation}
and%
\begin{align}
\psi_{0}\left(  \vec{r},t\right)   &  =N\int d^{3}r^{\prime}\exp\left[
\begin{array}
[c]{c}%
\begin{array}
[c]{c}%
\frac{im}{2\hbar t}\left(  z-z^{\prime}\right)  ^{2}+\frac{im\omega}{4\hbar
}\cot\left[  \frac{\omega t}{2}\right]  \left(  \left(  x-x^{\prime}\right)
^{2}+\left(  y-y^{\prime}\right)  ^{2}\right) \\
+\frac{im\omega}{2\hbar}\left(  xy^{\prime}-yx^{\prime}\right)
\end{array}
\\
-\frac{1}{2\sigma^{2}}\left(  x^{\prime2}+y^{\prime2}\right)  -\frac{1}%
{2L^{2}}z^{\prime2}+\frac{ip}{\hbar}z^{\prime}%
\end{array}
\right] \nonumber\\
&  =N\exp\left[  \frac{im}{2\hbar t}z^{2}+\frac{im\omega}{4\hbar}\cot\left[
\frac{\omega t}{2}\right]  \left(  x^{2}+y^{2}\right)  \right] \nonumber\\
&  \times\int d^{3}r^{\prime}\exp\left[  \alpha_{x}x^{\prime}+\alpha
_{y}y^{\prime}+\alpha_{z}z^{\prime}-\frac{1}{2\beta_{\rho}}\left(  x^{\prime
2}+y^{\prime2}\right)  -\frac{1}{2\beta_{z}}z^{\prime2}\right] \nonumber\\
&  =N\sqrt{\left(  2\pi\right)  ^{3}\beta_{\rho}^{~2}\beta_{z}}\nonumber\\
&  \times\exp\left[
\begin{array}
[c]{c}%
\left(  \frac{im\omega}{4\hbar}\cot\left[  \frac{\omega t}{2}\right]
-\frac{1}{2}\beta_{\rho}\left(  \frac{m\omega}{2\hbar\sin\left[  \frac{\omega
t}{2}\right]  }\right)  ^{2}\right)  \left(  x^{2}+y^{2}\right) \\
-\beta_{z}\left(  \frac{m}{\hbar t}\right)  ^{2}\left(  z-\frac{p}{m}t\right)
^{2}+\frac{im}{2\hbar t}z^{2}%
\end{array}
\right]  \label{3.14}%
\end{align}
where $N$ is the same as in (\ref{3.6}) but now%
\begin{align}
\beta_{\rho}  &  =\left(  \frac{1}{\sigma^{2}}-\frac{im\omega}{2\hbar}%
\cot\left[  \frac{\omega t}{2}\right]  \right)  ^{-1}\nonumber\\
\beta_{z}  &  =\left(  \frac{1}{L^{2}}-\frac{im}{\hbar t}\right)  \label{3.15}%
\end{align}
Because $\psi\left(  \vec{r},t\right)  $ depends on $x$ and $y$ only in the
combination $\rho^{2}=x^{2}+y^{2}$ it follows that the initial Gaussian wave
function chosen here does not pick up angular momentum as it propagates along
the magnetic field. In fact for propagation parallel to the magnetic field the
axial OAM of an eigenstate of $\mathbf{L}_{z}$ is conserved. This follows
directly from%
\begin{equation}
\left[  \mathbf{L}_{z}\mathbf{,H}\right]  =0 \label{3.16}%
\end{equation}
where again $\mathbf{H=}\left(  \overset{\rightarrow}{\mathbf{P}}-e\vec
{A}\left(  \overset{\rightarrow}{\mathbf{R}}\right)  \right)  ^{2}/2m$ and
$\mathbf{A}_{i}\mathbf{=-}\frac{B}{2}\epsilon_{ij}\mathbf{X}_{j}\mathbf{.}$
Indeed it can be shown that $\mathbf{H}=\frac{1}{2m}\overset{\rightarrow
}{\mathbf{P}}^{2}-\frac{eB}{2m}\mathbf{L}_{z}+\frac{e^{2}B^{2}}{2m}\left(
\mathbf{X}^{2}+\mathbf{Y}^{2}\right)  $ which obviously yields (\ref{3.16}).

\section{Conclusion}

Using the exact path integral solution for the propagator in a constant
magnetic field we have derived the evolution of a Gaussian wave function and
shown explicitly that the (non-radiatively corrected) gyromagnetic ratio
$g_{L}$ for OAM is unity. This must be the case since $g_{L}$ is a property of
the Hamiltonian and not of the wave function.

The results presented above a novel version of the Aharonov-Bohm
effect.\cite{Zee2} Consider a long thin solenoid aligned along the $z$
axis.\ Outside the solenoid (far from the ends) $\vec{A}$ varies as
$1/\rho=1/\sqrt{x^{2}+y^{2}}$ and so $\vec{B}$ is zero outside. Inside the
solenoid $\vec{A}$ varies as $\rho$ and so $\vec{B}$ is constant and nonzero.
A Gaussian wave function like those considered above carrying nozero OAM that
propagates along the $z$ axis has a node on the $z$ axis. In fact wave
functions carrying large values of OAM have a very large region around the $z$
axis where the wave function is effectively zero.\cite{Ben} As in the standard
Aharonov-Bohm experiment\cite{Zee2} this is a case where there is no overlap
between the wave function and the magnetic field. The wave function only
overlaps with the magnetic vector potential. Hence the presence of the
solenoid will cause a change in how the wave function propagates relative to
the no solenoid case. This effect will be predominantly a\ change in the focus
position of the wave function. Experimental verification of this would provide
yet another example of the fact $A_{\mu}$ is the fundamental quantity and not
$\vec{E}$ and $\vec{B}.$

\end{document}